# Minor Actinides Transmutation in Candidate Accident Tolerant Fuel−Claddings $U_3Si_2$-FeCrAl and $U_3Si_2$-SiC

Shengli Chen[1,2] and Cenxi Yuan[1,*]

[1] Sino-French Institute of Nuclear Engineering and Technology, Sun Yat-sen University, Zhuhai, Guangdong 519082, China
[2] CEA, Cadarache, DEN/DER/SPRC/LEPh, 13108 Saint Paul Les Durance, France
[*] Corresponding author. *E-mail address:* yuancx@mail.sysu.edu.cn

## Abstract

An advanced transmutation method is suggested that the long-lived Minor Actinides (MAs) in the spent fuel can be efficiently transmuted in the candidate Accident Tolerant Fuel (ATF). The transmutation of MAs is investigated through the Monte Carlo simulations in two potential fuel-claddings of ATF, $U_3Si_2$-FeCrAl and $U_3Si_2$-SiC. The critical loadings of MAs are determined through the Linear Reactivity Model (LRM) in order to keep the same reactivity as the current $UO_2$-zircaloy system at the End of Cycle (EOC). In all cases, excellent transmutation efficiencies are found for the most important three MAs, $^{237}$Np, $^{241}$Am, and $^{243}$Am, of which the total transmutation rates are around 60%, 90%, and 60%, respectively. If only the longest-lived isotope $^{237}$Np is considered, one $U_3Si_2$-SiC assembly can transmute $^{237}$Np from six normal assemblies. The loading of MAs has little influences on the neutronic properties, such as the power distributions inside an assembly and inside a fuel rod. The transmutation of MAs in the ATF assembly is shown to be more efficient and safe comparing with the normal assembly, while other important properties are kept, such as the cycle length and the power distribution.

## 1. Introduction

The Accident Tolerant Fuel (ATF) has been developed for the advanced nuclear fuel and cladding options after the Fukushima nuclear disaster in 2011. The US DOE NE Advanced Fuels Campaign has proposed many candidate ATFs and claddings [1]. As explained in Ref. [2], the FeCrAl cladding has better accident tolerant performance than the current zircaloy cladding. FeCrAl has also been studied in the CANDU reactor [3]. However, FeCrAl has about 10 times larger thermal neutron absorption cross sections than that of zircaloy [4]. In order to composite its large neutron capture cross section, the $U_3Si_2$ fuel [5] has been proposed. $U_3Si_2$ is also a candidate ATF of Westinghouse [6], [7]. The stability of the $U_3Si_2$ fuel under coolant conditions has also been investigated [8]. The neutronic studies show that the $U_3Si_2$-FeCrAl fuel-cladding combination can be a potential ATF combination [2], [9].

SiC is another excellent candidate ATF cladding material due to its exceptional oxidation resistance in high temperature steam environment [10]. Its maximum service temperature is 900°C, while that of the zirconium alloy is only 400°C [11]. Its stability at high neutron fluence has also been proved [12]. In addition, the thermal neutron absorption cross section of SiC is about 10 times smaller than that of the zircaloy [11]. SiC is also candidate cladding material for generation IV reactors [11].

The management of long-lived radioactive products in the spent nuclear fuel is one of the most difficult issues in nuclear engineering. Actinides have the most important contribution on radioactivity in the spent fuel. The closed nuclear fuel cycle and the transmutation of some long-lived radioactive nuclides have been proposed as an alternative concept for solving the problem. Uranium and plutonium isotopes are extracted from the spent fuel in the closed fuel cycle. The extracted plutonium can be used to fabricate uranium-plutonium Mixed Oxide (MOX) nuclear fuel. The treatment of long-lived Minor Actinides (MAs) becomes one of the most urgent issues to reduce the long-lived radiotoxic isotopes in the spent fuel. Accordingly, the study of the transmutation of MAs is a significant work for the post-processing of the spent fuel. The transmutation of MAs has been largely studied in different types of reactors, such as typical uranium fueled [13], [14] and MOX fueled [15], [16] Light Water Reactors (LWRs), Fast neutron Reactors (FRs) [17], [18], and Accelerator Driven Sub-critical reactor (ADS) [19]. Comparing with the transmutation in the fast reactors or other advanced reactors, the transmutation in Pressurized Water Reactors (PWRs) is important because 65% operational nuclear reactors are PWRs in the world in 2017 [20].

Due to the high uranium density in the candidate ATF $U_3Si_2$, the previous study shows that 4.9% enriched $U_3Si_2$ and 350 μm thickness FeCrAl cladding has higher reactivity at the End of Cycle (EOC) than the current $UO_2$-zircaloy system [2]. On the other hand, because of the competition of absorbing neutrons between MAs and fissile isotopes, the transmutation of MAs in reactors has a negative contribution to reactivity. Therefore, the present work proposes the transmutation of MAs in the 4.9% enriched $U_3Si_2$ fuel and 350 μm thickness FeCrAl cladding combination.

As explained previously, SiC is also a potential accident tolerant cladding. Contrary to FeCrAl, the thermal neutron absorption cross section of SiC is much smaller than that of zircaloy. The present work performs on the transmutation of MAs in Candidate ATF $U_3Si_2$ combined with SiC cladding.

Due to the spatial self-shielding in Pressurized Power Reactors (PWR), the periphery phenomenon in a fuel rod is evident [21], [22]. As new fuel-cladding combinations, studies on their radial distribution of physical properties are of importance. For example, the radial power distribution can give feedback to neutronic calculations and multi-physics coupling study [23]. The investigation on the radial distribution of the isotopic concentrations of major actinides can provide useful information to the possible prolongation of the fuel life. The study performed on the radial distribution of the concentrations of MAs is a method to investigate the local transmutation efficiency.

The beginning of Section 2 gives a brief presentation of the fuel characters and the simulation method. The method to determine the percentage of MAs in the $U_3Si_2$ fuel and the methods to evaluate the transmutation efficiency of MAs are explained in this section. At the end of this section, the empirical formula obtained in the previous study is used to describe the radial distribution of power and isotopic concentration. Section 3 shows the results and corresponding discussions, including the percentage of MA in each fuel-cladding combination based on the cycle length analysis, the efficiencies of transmutation, and the radial distribution of power and isotopic concentration. The last section is summary and conclusions.

# 2. Method

## 2.1 Simulation methods

A typical 17×17 PWR fuel assembly is used in the present study. Figure 1 illustrates the configuration of the lattice of the assembly. The assembly is constituted by 264 fuel rods and 25 tubes, including 24 guide tubes and a central instrumentation tube. In the present study, these tubes are full of the moderator because of no insertion of control rods or instrument. The key parameters of the fuel assembly are given in Table 1. More parameters of this kind of fuel assembly can be found in Ref. [2].

In general, the decay half-life of MA is very long. The half-lives of the five MAs considered in the present work are shown in Table 2. Due to the long half-life of $^{237}$Np and its large quantity in the spent fuels, the transmutation of $^{237}$Np is of large importance for the post-processing of the spent fuel. In addition, in order to transmute other long-lived MAs, it is interesting to study the mixed long-lived MAs. The percentage of each MA, shown in Table 2, corresponds to the relative percentage of these 5 MAs in the 10 years cooled spent fuel unloaded from a 3 GW thermal power reactor at 30 GWd/t burnup level [24].

Table 1. Key parameters of the fuel assembly design

| Property | Unit | Value |
|---|---|---|
| FeCrAl composition | wt% | Fe:Cr:Al = 75:20:5 |
| SiC composition | at% | Si:C=1:1 |
| Gap thickness | μm | 82.55 |
| Cladding thickness | μm | 350 & 571.5 |
| Cladding Outer Radius | mm | 4.7498 |
| Fuel enrichment | % | 4.9 |
| Fuel temperature | K | 900 |
| Pitch to fuel rod outer diameter | - | 1.326 |
| Fuel density | g/cm$^3$ | 11.57 |
| Coolant density | g/cm$^3$ | 0.7119 |
| Coolant temperature | K | 580 |
| Cladding density | g/cm$^3$ | FeCrAl: 7.10<br>SiC: 2.58 |
| Cladding temperature | K | 580 |
| Boron concentration | ppm | 630 |

Table 2. Decay half-life of MAs and the relative mass percentage in the 10 years cooled spent fuel

| Isotope | $^{237}$Np | $^{241}$Am | $^{243}$Am | $^{244}$Cm | $^{245}$Cm |
|---|---|---|---|---|---|
| Half-life (year) [25] | 2.144×10$^6$ | 432.6 | 7364 | 18.1 | 8324 |
| Percentage (%) [24] | 41.80 | 47.86 | 8.62 | 1.63 | 0.09 |

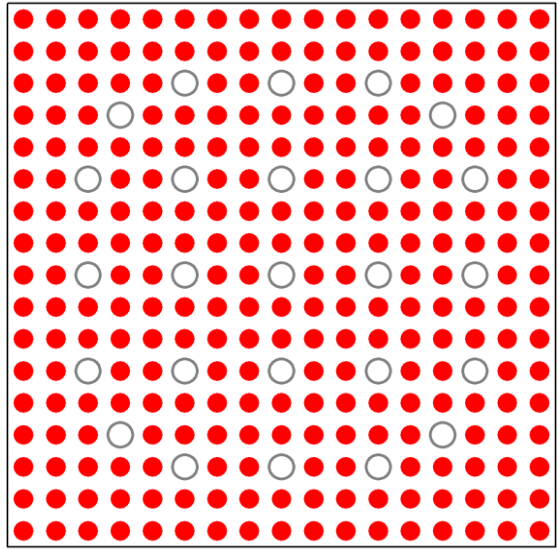

Figure 1. 17×17 PWR lattice configuration

The uranium enrichment of the $U_3Si_2$ fuel is the same as current Westinghouse PWR, such as 4.9%. FeCrAl and SiC claddings are used for MAs loaded $U_3Si_2$ transmutation fuel study. In Light Water Reactor (LWR), 350 μm stainless steel cladding was used [26]. This thickness is utilized in the present work. The same thickness is used for the ceramic SiC cladding, referred to SiC-1 hereinafter. Due to the lack of the applied experience of the SiC cladding, the cladding thickness 571.5 μm of the current zircaloy cladding is also used for SiC and it is hereinafter referred as SiC-2.

In general, the Monte Carlo method is reliable for the neutronic calculations of new fuel-cladding combinations. One important reason is that the nuclear data libraries with point-wise cross sections are essentially suitable for various neutron spectrums and self-shielding effects. The deterministic method is also available and maybe more suitable for the sub-layer power calculations if it is well benchmarked for the neutron spectrum and the self-shielding effect of a new system. In the present work, the main purpose is the comparison of the transmutation efficiencies among various fuel-cladding combinations loading different types and amounts of MAs. It is better to use Monte Carlo based code to have general comparisons among different schemes.

All the neutronic simulations in the present study are performed with the Monte Carlo based code RMC [27]. RMC is a 3D stochastic neutron transport code developed by Tsinghua University. The continuous-energy pointwise cross sections are used for the different materials. The present investigations focus on the calculation of the infinity multiplication factor $k_{inf}$, the nuclear power, and isotopic concentrations for different cases by using the ENDF/B-VII.0 library [28]. In nuclear physics, the fundamental properties of the atomic nuclei can be solved through the many-body Schrodinger Equation [29], [30], of which the calculation can be simplified through the Monte Carlo method. The present simulations take 1030 batches (with 30 first discarded batches) and 10 000 neutrons per batch. The depletion calculations are based on 3.3 Effective Full Power Days (EFPDs), 13.3 EFPDs, 16.7 EFPDs for the 3 first steps and each 33.3 EFPDs from the forth step to the end. Each step is divided by 10 sub-steps to calculate isotopic concentrations for the next burnup level with Bateman equation by supposing constant isotopic concentrations in each sub-step.

**2.2 Transmutation studies**

The criterion to determine the percentage of MAs is the equivalent reactivity at the End of Cycle (EOC). The formula to calculate the average eigenvalue difference is [31]:

$$\Delta k_{core} = \frac{\sum_b \Delta k_{inf,b}(e_b) P_b V_b}{\sum_b P_b V_b}, \quad (1)$$

where $\Delta k_{inf,b}$ is the difference of infinity multiplication factor $k_{inf}$ between the investigated fuel-cladding combination and the reference case for batch $b$ as a function of exposure $e_b$. The fuel exposure (in EFPD) at the EOC from Table 3 are used to quantify the level of exposure for each batch in a typical Westinghouse PWR reactor. The power weighting factor $P_b$ is the approximative contribution of each batch to the power distribution in the core. $V_b$ refers to the number of assemblies per batch. The positive (negative) value of $\Delta k_{core}$ at the EOC signifies that the studied fuel-cladding combination has a longer (shorter) cycle length than the reference case. In order to avoid the reduction of cycle length, the MAs loaded fuel-cladding combinations should satisfy $\Delta k_{core} \geq 0$. The condition to determine the critical MA loading by keeping the cycle length is $\Delta k_{core} = 0$.

Table 3. Distribution of population and power per fuel cycle batch in a typical Westinghouse PWR [32]

| Batch | Number of assemblies | Core fraction vol% ($V_b$) | Relative assembly power ($P_b$) | EFPDs achieved at EOC ($e_b$) |
|---|---|---|---|---|
| 1 | 73 | 38% | 1.25 | 627 |
| 2 | 68 | 35% | 1.19 | 1221 |
| 3 | 52 | 27% | 0.40 | 1420 |
| Total | 193 | 100% | - | - |

To study the transmutation efficiency, the following three quantities are defined [15]: the net transmutation rate $R_N$ (Eq. (2)); the total transmutation rate $R_T$ (Eq. (3)); and the equivalent natural decay time $Te$ (Eq. (4)). $R_N$ defines the percentage of transmuted MA to loading quantity. $R_T$ is the transmutation rate taking the production of MA in normal assembly into account. $Te$ represents the equivalent decay time to achieve the $R_T$ by natural decay. $Te$ measures the efficiency of transmutation on the reduction of the time duration of storage, while the transmutation rates account the reduction of the quantity of MA. By the above definitions,

$$R_N = (C_0 - C_2)/C_0 \times 100\%, \quad (2)$$

$$R_T = (C_0 - C_2 + C_1)/C_0 \times 100\%, \quad (3)$$

$$Te = -T_{1/2} \ln(1-R_T)/\ln 2, \quad (4)$$

where $C_0$ stands for the initial concentration of the MA added in the fuel, $C_1$ and $C_2$ are concentrations of the MA after depletion in the none MA loaded case and MAs loaded cases, respectively.

We remark that no MA loading fuel has longer cycle length than transmutation fuel if all conditions are the same except the percentage of MAs. The concentration $C_l$ is thus an approximation, which is calculated in no MA loading fuel assembly. Ref. [2] shows that the 4.58% enriched $U_3Si_2$ fuel and FeCrAl cladding with 350 μm thickness combination has the same cycle length as the reference case. Table 4 shows the process to produce the above five MAs in the nuclear fuel. More details of the discussion about the reactions concerning the actinides in nuclear fuels can be found in Ref. [33]. The maximum deviation of $C_l$ should be the concentration of $^{237}Np$ when fuel enrichment is changed. Due to the long decay half-life of $^{241}Am$, $^{237}Np$ mainly comes from $^{235}U$. If the 4.58% fuel enrichment is used, $C_l$ should be about 6% lower than the 4.9% enriched fuel. Other four MAs are from a series of reactions of $^{238}U$, so the influence of the fuel enrichment on their concentrations can be neglected.

Table 4. Process of the production of the above five MAs in nuclear fuel

| MA | Production process |
|---|---|
| $^{237}Np$ | $^{235}U \xrightarrow{n,\gamma} {^{236}U} \xrightarrow{n,\gamma} {^{237}U} \xrightarrow{\beta^-} {^{237}Np}$ or $^{241}Am \xrightarrow{\alpha} {^{237}Np}$ |
| $^{241}Am$ | $^{238}U \xrightarrow{n,\gamma} {^{239}U} \xrightarrow{\beta^-} {^{237}Np} \xrightarrow{n,\gamma} {^{239}Pu} \xrightarrow{n,\gamma} {^{240}Pu} \xrightarrow{n,\gamma} {^{241}Pu} \xrightarrow{\beta^-} {^{241}Am}$ |
| $^{243}Am$ | $^{238}U \xrightarrow{n,\gamma} {^{239}Pu} \xrightarrow{n,\gamma} {^{240}Pu} \xrightarrow{n,\gamma} {^{241}Pu} \xrightarrow{n,\gamma} {^{242}Pu} \xrightarrow{n,\gamma} {^{243}Pu} \xrightarrow{\beta^-} {^{243}Am}$ |
| $^{244}Cm$ | $^{238}U \xrightarrow{n,\gamma} {^{239}Pu} \xrightarrow{n,\gamma} {^{240}Pu} \xrightarrow{n,\gamma} {^{241}Pu} \xrightarrow{n,\gamma} {^{242}Pu} \xrightarrow{n,\gamma} {^{243}Pu} \xrightarrow{\beta^-} {^{243}Am} \xrightarrow{n,\gamma} {^{244}Am} \xrightarrow{\beta^-} {^{244}Cm}$ |
| $^{245}Cm$ | $^{238}U \xrightarrow{n,\gamma} {^{239}Pu} \xrightarrow{n,\gamma} {^{240}Pu} \xrightarrow{n,\gamma} {^{241}Pu} \xrightarrow{n,\gamma} {^{242}Pu} \xrightarrow{n,\gamma} {^{243}Pu} \xrightarrow{\beta^-} {^{243}Am} \xrightarrow{n,\gamma} {^{244}Am} \xrightarrow{\beta^-} {^{244}Cm} \xrightarrow{n,\gamma} {^{245}Cm}$ |

## 2.3 Radial properties

With the above transmutation fuel-cladding combinations, the present work investigates their radial distributions of the power and the isotopic concentration in a fuel pellet. The geometry of the fuel rod is shown in Figure 2. The fuel region is divided into 9 concentric rings for the calculations of the radial distribution of the power and the isotopic concentrations, while the gap and cladding are located outside.

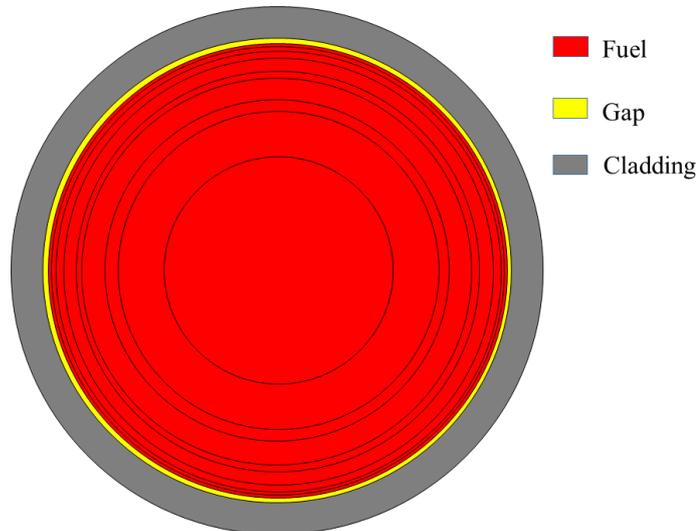

Figure 2. Radial profile of a fuel-cladding system with fuel region divided into 9 rings

To describe the relative power distribution as a function of burnup, the previous work [34] uses the second order polynomial:
$$f(x,s) = a(x)s^2 + b(x)s + c(x), \quad (5)$$
where $x = r/r_0$ with $r_0$ is the radius of fuel, $s$ represents the fuel equivalent full power depletion time, and the coefficients of polynomial depend on the position $x$. This form can be used for both the radial power distribution and the major actinide for six candidate ATF fuel-cladding combinations. Eq. (5) is also available to fit the isotropic concentrations of $^{235}$U, $^{238}$U, $^{239}$Pu, and $^{241}$Pu [34].

## 3. Results and Discussion

### 3.1 Critical loadings of MAs

Taking the current UO$_2$-Zircaloy fuel-cladding system as the reference, the differences on the reactivity induced by the different MA loading percentages (*lp*) can be obtained by using the Linear Reactivity Model (LRM). The infinity multiplication factor $k_{inf}$ for each *lp* is calculated with the Monte Carlo simulation. We note $\Delta k_{core}$ for the difference of the multiplication factor in the core at EOC. $\Delta k_{core}$ values are given in Table 5 and Table 6 for different MA loading cases. The loading percentages of MAs with $\Delta k_{core}$ far from zero are not analyzed because the objective is to find the critical loading concentrations of MAs in the different fuel-cladding combinations. The reactivity decreases with the loading of MAs because of the competition of neutrons between MAs and fissile isotopes. For the same loaded amount of MAs, the reactivity decreases with the thermal neutron absorption ability of the cladding (FeCrAl > SiC-2 > SiC-1) due to the competition of absorbing thermal neutrons between fuel and cladding.

In order to determine the relationship between $\Delta k_{core}$ and *lp*, the linear function is supposed:
$$\Delta k_{core} = a + b \cdot lp. \quad (7)$$
The coefficients and the coefficient of determination are shown in Table 7 for the six cases studied in the present work. The near to unity values of the coefficient of determination point out that the assumption of linear relationship between $\Delta k_{core}$ and *lp* is validated. The values of *a* are always positive because the three chosen fuel-cladding combinations have longer cycle length than the reference fuel-cladding system without loading MAs. It is also the reason that the transmutations of MAs in these combinations are proposed in the present work. The negative values of *b* are due to the competition of neutrons between fissile nuclei and MAs. The physical significance of *b* is the negative contribution on reactivity by loading 1% MAs.

Table 5. $\Delta k_{core}$ between transmutation core with different loading percentages (*lp*) of mixed MAs/$^{237}$Np and the reference case for the FeCrAl cladding

| *lp* (%) | 0.0 | 0.3 | 0.32 | 0.5 |
|---|---|---|---|---|
| MAs | 0.0229 | 0.0006 | -0.0010 | -0.0135 |
| $^{237}$Np | 0.0229 | 0.0008 | -0.0008 | -0.0136 |

Table 6. $\Delta k_{core}$ between transmutation core with different loading percentages (*lp*) of mixed MAs/$^{237}$Np and the reference case for SiC claddings of 350 μm (SiC-1) and 571.5 μm (SiC-2) thickness

| *lp* (%) | 0.0 | 0.3 | 0.5 | 0.7 | 0.8 | 1.0 |
|---|---|---|---|---|---|---|
| SiC-1/MAs | - | - | 0.0184 | 0.0031 | -0.0034 | -0.0179 |
| SiC-1/$^{237}$Np | - | - | 0.0194 | 0.0052 | -0.0018 | -0.0150 |
| SiC-2/MAs | 0.0432 | 0.0209 | 0.0072 | -0.0072 | - | - |
| SiC-2/$^{237}$Np | 0.0432 | 0.0204 | 0.0060 | -0.0066 | - | - |

Table 7. Linear fitting results for the FeCrAl cladding, 350 μm (SiC-1) and 571.5 μm (SiC-2) thickness SiC cladding with mixed MAs and $^{237}$Np loading. *a*, *b* are constants defined in Eq. (7), $R^2$ is the coefficient of determination of fitting, and *c* is the critical loading of MAs to keep the cycle length. The last column lists the corresponding $\Delta k_{core}$ between transmutation core with critical mixed MAs and $^{237}$Np loading *c* and the reference case.

| Case | *a* | | *b* | | $R^2$ | *c* (%) | $\Delta k_{core}$ |
|---|---|---|---|---|---|---|---|
| FeCrAl/MAs | 0.02268 | 1.31% | -0.07305 | 1.24% | 0.9994 | 0.31 | -0.0001 |
| FeCrAl/$^{237}$Np | 0.02277 | 0.83% | -0.07311 | 0.79% | 0.9998 | 0.31 | 0.0000 |
| SiC-1/MAs | 0.05431 | 1.82% | -0.07235 | 1.77% | 0.9991 | 0.75 | -0.0002 |
| SiC-1/$^{237}$Np | 0.05363 | 1.56% | -0.06889 | 1.57% | 0.9993 | 0.78 | -0.0005 |
| SiC-2/MAs | 0.04294 | 0.79% | -0.07176 | 1.04% | 0.9997 | 0.60 | -0.0001 |
| SiC-2/$^{237}$Np | 0.04252 | 2.22% | -0.07141 | 2.91% | 0.9975 | 0.60 | -0.0005 |

The penultimate column in Table 7 presents the critical MA loading percentages (*c*), which means the same reactivity for the transmutation and reference core at EOC ($\Delta k_{core} = 0$). The critical MA loading decreases with the stronger thermal neutron absorption ability of the cladding because of the competition of absorbing thermal neutrons between fuel and cladding. The relative production of MAs at the EOL in the reference case are 0.0740%, 0.0057%, and 0.0207% for $^{237}$Np, $^{241}$Am, and $^{243}$Am, respectively. In the pure $^{237}$Np loading cases, the scenarios of FeCrAl cladding, SiC-1 cladding, and SiC-2 cladding transmute respectively 4.2, 10.5, and 8.1 times of $^{237}$Np produced in the standard fuel at the EOL. If the larger volume of nuclear fuel due to the thinner cladding is accounted for the FeCrAl cladding and SiC-1 cladding cases, the factors of transmuted $^{237}$Np are 4.7 and 11.7, respectively. In the mixed MAs loading cases, less $^{237}$Np is transmuted due to the loading of other MAs. However, the transmutation of $^{241}$Am is quite significant. 26, 63, and 54 times of $^{241}$Am produced in the standard spent fuel are transmuted. The factors of 29 and 70 are found if the volume is considered for the FeCrAl cladding and SiC-1 cladding, respectively.

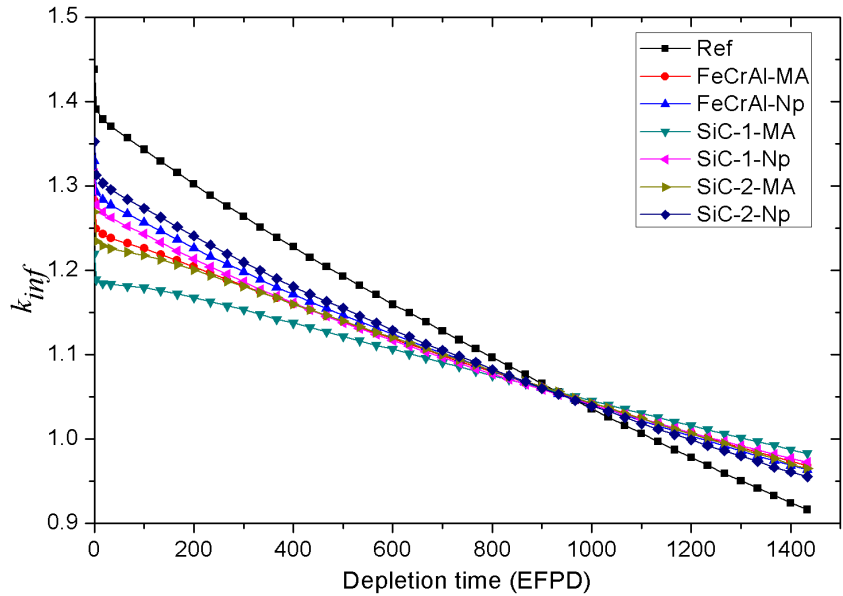

Figure 3. Infinity multiplication factor $k_{inf}$ versus effective depletion time without considering impact of reactor stops

The Monte Carlo simulations are used to verify whether the same cycle lengths are obtained with the corresponding $c$ values in Table 7. $k_{inf}$ for the reference case and the above six transmutation cases are shown in Figure 3 as a function of effective full power depletion time. The calculations are performed without considering impact of reactor stops, despite several reactor cycles are involved. The variations of $k_{inf}$ for the transmutation scenarios are smoother than the reference case. At the BOL, the reactivity of transmutation combinations is lower due to the competition of neutrons between MAs and fissile nuclei. At high fuel exposure, since the MAs are transmuted to fissile nuclei, higher reactivity is found in the transmutation cases. The flatten of $k_{inf}$ versus fuel exposure is more evident in mixed MAs loading scenarios than the pure $^{237}$Np loading because the thermal neutron capture cross section of $^{241}$Am is larger than that of $^{237}$Np (see Table 15). The variation of $k_{inf}$ shows that the MAs can be considered as burnable poison. The corresponding $\Delta k_{core}$ value for each case is given in the last column in Table 7. The Monte Carlo calculation shows the high accuracy of the critical MA loading percentages obtained from the linear fitting.

As expected, the percentage of loaded MAs in the SiC cladding case is larger than that in the FeCrAl cladding case because of the much lower thermal neutron absorption cross section of SiC. The good neutron economy of the SiC cladding is evident. The mixed MAs loading and $^{237}$Np loading have different influence on $\Delta k_{core}$ due to the different behaviors between $^{237}$Np and other MAs. However, the difference on $\Delta k_{core}$ is not obvious in the FeCrAl cladding combinations and in the SiC-2 cladding cases. For SiC-1 cladding assemblies, the difference is more evident than the FeCrAl and SiC-2 cases. This should come from the higher concentrations of MAs, which enhances the difference of neutronic behavior between $^{237}$Np and other MAs. Anyway, this difference is generally very small, which is less than 4% among different loadings.

The transmutation fuel assemblies have lower reactivity than the current assembly at the low burnup level. One of the most important reasons is the neutron absorption reaction of MAs. The opposite effect is observed at the high burnup level because the neutron capture reaction of fertile MAs produces the fissile isotopes, such as $^{242}$Am, $^{242m}$Am and $^{244}$Am with huge fission cross sections, especially at thermal region [28], [35]. The thermal neutron capture cross section of $^{241}$Am is larger than that of $^{237}$Np. The flattening of the reactivity versus burnup in the mixed MAs loading fuels is thus more evident.

### 3.2 Transmutation efficiencies of MAs

The transmutation efficiency of each MA in each case is calculated at the End of Life (EOL, equivalent to 1420 EFPDs) of the fuel assemblies. Table 8 and Table 9 show the transmutation efficiency in the pure $^{237}$Np loading cases and in the mixed MAs loading cases, respectively. Table 10 gives the sum of the radioactivity over the initially loaded five MAs ($A_1$) in each fuel assembly at the Beginning of Life (BOL) and at the EOL. $A_2$ ($A_3$) is the sum of the radioactivity over the above MAs of which the half-lives are longer than 5 000 (100 000) years. Among the above five MAs, only the $^{237}$Np has the half-life longer than ten thousand years, so $A_3$ is equivalent to the radioactivity of $^{237}$Np.

Table 8. Transmutation efficiency in the $^{237}$Np loading cases with FeCrAl cladding and 350 μm (SiC-1) and 571.5 μm (SiC-2) thickness SiC cladding. $R_N$, $R_T$, and $Te$ are the net transmutation rate, total transmutation rate, and the equivalent natural decay time, respectively.

| Cladding | $R_N$ (%) | $R_T$ (%) | $Te$ ($10^6$ year) |
|---|---|---|---|
| FeCrAl | 41.3 | 63.1 | 3.09 |
| SiC-1 | 52.9 | 61.4 | 2.95 |
| SiC-2 | 53.6 | 65.4 | 3.28 |

Table 9. Transmutation efficiency in the mixed MAs loading cases with FeCrAl cladding and 350 μm (SiC-1) and 571.5 μm (SiC-2) thickness SiC cladding. $R_N$, $R_T$, and $Te$ are the net transmutation rate, total transmutation rate, and the equivalent natural decay time, respectively.

| Cladding | MA | $R_N$ (%) | $R_T$ (%) | $Te$ (year) |
|---|---|---|---|---|
| FeCrAl | $^{237}$Np | 11.6 | 62.7 | 3.05×10$^6$ |
|  | $^{241}$Am | 87.8 | 92.1 | 1.59×10$^3$ |
|  | $^{243}$Am | 25.9 | 62.0 | 1.03×10$^4$ |
| SiC-1 | $^{237}$Np | 41.2 | 61.5 | 2.95×10$^6$ |
|  | $^{241}$Am | 89.4 | 91.1 | 1.51×10$^3$ |
|  | $^{243}$Am | 39.4 | 56.9 | 8.93×10$^3$ |
| SiC-2 | $^{237}$Np | 38.4 | 65.2 | 3.26×10$^6$ |
|  | $^{241}$Am | 91.6 | 93.7 | 1.73×10$^3$ |
|  | $^{243}$Am | 32.4 | 60.1 | 9.76×10$^3$ |

According to the brief analysis in Section 2, the relative uncertainty of $C_I$ due to the choice of the reference case should be less than 6%. On the other hand, $C_I$ is proportional to the difference of $R_T$ to $R_N$ (Eq. (3) subtracts Eq. (2)). Because $R_N$ is independent to $C_I$, the uncertainty of $R_T$ due to the indetermination of $C_I$ ($dC_I$) is expressed as:

$$dR_T = dC_I/C_I \times (R_T - R_N). \qquad (8)$$

By using Eq. (8) and $R_T$ and $R_N$ in Table 8 and Table 9, one can conclude that the approximation of $C_I$ used in the present work leads around 1% error on total transmutation rate, except for the 3% error in the 0.31% mixed MAs loading case with the FeCrAl cladding. The above uncertainties are based on the rough comparison with the 4.58% enriched fuel. If the 4.9% enriched fuel with thicker cladding (to keep the same cycle length) is used as the normal case, the deviation of the total transmutation rate would be even less. By consequent, the $C_I$ used in the present work has a very limited influence on the calculations of the total transmutation rate.

For $^{237}$Np, the total transmutation efficiency (given in Table 8) is higher than 60% in the six cases. It is equivalent to $3\times10^6$ years of natural decay. Therefore, comparing with the same reduction through natural decay in deep geological repositories, the transmutation is an excellent method to reduce the quantity of $^{237}$Np. As shown in Table 9, both the net transmutation rate and the total efficiency are about 90% for $^{241}$Am in the three mixed MAs loading cases. The excellent transmutation rate is due to its large thermal neutron capture cross section. Therefore, the transmutation proposed in the present work is a very efficient way to reduce the quantity of $^{241}$Am. The concentrations of $^{237}$Np and $^{241}$Am are about 90% among the above five MAs in the spent fuel. $^{243}$Am is the dominant isotope (8.6%) in the rest of 10% MAs. As given in Table 9, 26% to 39% of the initial loading $^{243}$Am is transmuted in the above three transmutation cores. The total efficiency is about 60% for the three cases. Therefore, the transmutation of $^{243}$Am has an evident effect on the reduction of MAs.

Table 10. Radioactivity in the fuels with FeCrAl cladding and 350 μm (SiC-1) and 571.5 μm (SiC-2) thickness SiC cladding for no MA loading, mixed MAs loading, and $^{237}$Np loading cases at the BOL and EOL. $A_1$ and $A_2$ ($A_3$) are the sum of radioactivity over the above five MAs and that over the above five MAs with the half-lives longer than 5 000 (100 000) years.

| Cladding | Loading | Fuel status | $A_1$ ($10^{11}$ Bq/kg) | $A_2$ ($10^8$ Bq/kg) | $A_3$ ($10^7$ Bq/kg) |
|---|---|---|---|---|---|
| FeCrAl | No MA | EOL | 1.19 | 7.55 | 1.52 |
| | MAs | BOL | 3.08 | 22.0 | 2.97 |
| | | EOL | 7.40 | 17.4 | 2.63 |
| | $^{237}$Np | BOL | $6.95\times10^{-4}$ | 0.69 | 6.95 |
| | | EOL | 1.19 | 7.84 | 4.08 |
| SiC-1 | No MA | EOL | 1.13 | 7.31 | 1.49 |
| | MAs | BOL | 6.84 | 44.8 | 7.32 |
| | | EOL | 14.0 | 29.1 | 4.31 |
| | $^{237}$Np | BOL | $1.75\times10^{-3}$ | 1.75 | 17.5 |
| | | EOL | 1.14 | 8.08 | 8.24 |
| SiC-2 | No MA | EOL | 1.59 | 9.64 | 1.59 |
| | MAs | BOL | 6.03 | 37.4 | 5.95 |
| | | EOL | 12.63 | 26.5 | 3.66 |
| | $^{237}$Np | BOL | $1.35\times10^{-3}$ | 1.35 | 13.5 |
| | | EOL | 1.60 | 10.2 | 6.25 |

$^{244}$Cm and $^{245}$Cm are not discussed in this section because of their very low initial loading concentration. In fact, the half-life of $^{244}$Cm is 18 years. Such short half-life signifies that the quantity of $^{244}$Cm will reduce to 3% in 90 years. The transmutation rate of $^{244}$Cm is not very important. The concentration of $^{245}$Cm is less than 0.1% among the above five MAs and it is a fissile isotope due to its even-odd proton-neutron property. Their effects will be discussed together with all MAs in Section 3.3.

$A_1$ is the total radioactivity of the above discussed five MAs. It describes the radiation dose induced by the five MAs. The column about $A_1$ in Table 10 shows that the radioactivity at EOL is higher than that at BOL for all cases. This is mainly due to the production of $^{244}$Cm, of which the half-life is much shorter than other four MAs. $A_2$ measures the radioactivity of long-lived MAs. The smaller values of $A_2$ at the EOL than the values at the BOL for the mixed MAs loading cases show the reduction of radiotoxicity induced by long-lived MAs through the transmutation. $A_3$ corresponds to very long-live MAs, of which the half-life is longer than 100 000 years. The comparison of values of $A_3$ at the BOL and EOL reveal the transmutation efficiency for isotopes with half-life longer than 100 000 years. Actually, only $^{237}$Np has such long half-life among the above five MAs. The reduction of $^{237}$Np is of the most significance because it needs to be stored underground for some million years. If only $^{237}$Np is considered in the transmutation, the results in Table 10 show that one SiC-1 assembly can transmute $^{237}$Np produced from six assemblies in a normal PWR while other properties are kept, including the cycle length, $A_1$, $A_2$, and radial properties discussed later.

**3.3 Buildup in curium and californium**

The transmutation is shown efficient to reduce the quantity of MAs. One of the drawbacks of transmutation is the production of heavier nuclei through several radiative capture reactions and potential β decays. An inventory of other residual MAs is performed to show their influence on the transmutation. The present work investigates the transmutation of long-lived MAs, the short-lived MAs, such as $^{242}$Am and $^{241,242}$Cm, are not of interest in this kind of studies. Table 11 shows the predominant decay modes and corresponding half-lives of MAs. It is noticeable that all isotopes of Bk and Es have relatively short half-lives. Only the MAs of which the half-lives are longer than 1 year (boldfaced in Table 11) are considered to evaluate the buildup in other residual MAs due to the transmutation.

The atomic concentrations at the EOL of MAs, of which the half-lives are longer than 1 year, are given in Table 12. The loading of pure $^{237}$Np has negative contributions on concentrations of Cm and Cf. In fact, the loading of $^{237}$Np decreases the initial concentrations of $^{238}$U, which is the dominant precursor of MAs other than $^{237}$Np. The increase of Pu by loading $^{237}$Np is evident due to serval neutron capture reactions and a β decay (such as shown in Table 11, $^{239}$Np decays to $^{239}$Pu). As explained in the Introduction, Pu is recycled in closed fuel cycle and not considered in the present studies. Mixed MAs loading leads to the higher concentrations of Cm and Cf because of the capture reactions and β decays of loaded MAs. After $^{237}$Np and $^{241,243}$Am, the two most concentrated isotopes are $^{244}$Cm and $^{245}$Cm. The high concentrations of $^{244}$Cm and $^{245}$Cm are due to both the initial loading and transmutation of americium.

Table 11. Main decay mode and half-life of Am, Cm, Bk, Cf, and Es isotopes [25]

| Isotope | Half-life | Decay mode | Product | Half-life | Decay mode | Product |
|---|---|---|---|---|---|---|
| **$^{241}$Am** | **433 d** | α | $^{237}$Np | **2.14×10$^6$ y** | α | $^{234}$Pa |
| $^{242}$Am | 16 h | β/β$^+$ | $^{242}$Cm/$^{242}$Pu | | | |
| **$^{243}$Am** | **7364 y** | α | $^{239}$Np | 2 d | β | $^{239}$Pu |
| $^{241}$Cm | 33 d | β$^+$ | $^{241}$Am | | | |
| $^{242}$Cm | 163 d | α | $^{238}$Pu | | | |
| **$^{243}$Cm** | **21 y** | α | $^{239}$Pu | | | |
| **$^{244}$Cm** | **18 y** | α | $^{240}$Pu | | | |
| **$^{245}$Cm** | **8423 y** | α | $^{241}$Pu | | | |
| **$^{246}$Cm** | **4706 y** | α | $^{242}$Pu | | | |
| **$^{247}$Cm** | **1.56×10$^7$ y** | α | $^{243}$Pu | | | |
| **$^{248}$Cm** | **3.48×10$^5$ y** | α | $^{244}$Pu | | | |
| $^{249}$Cm | 64 m | β | $^{249}$Bk | | | |
| **$^{250}$Cm** | **3.48×10$^5$ y** | **SF** | - | | | |
| $^{251}$Cm | 17 m | β | $^{251}$Bk | | | |
| $^{249}$Bk | 330 d | β | $^{249}$Cf | | | |
| $^{250}$Bk | 330 h | β | $^{250}$Cf | | | |
| $^{251}$Bk | 56 m | β | $^{251}$Cf | | | |
| **$^{249}$Cf** | **351 y** | α | $^{245}$Cm | | | |
| **$^{250}$Cf** | **13 y** | α | $^{246}$Cm | | | |
| **$^{251}$Cf** | **898 y** | α | $^{247}$Cm | | | |
| **$^{252}$Cf** | **2.6 y** | α | $^{248}$Cm | | | |
| $^{253}$Cf | 18 d | β | $^{253}$Es | | | |
| $^{254}$Cf | 61 d | SF | - | | | |
| $^{255}$Cf | 18 m | β$^-$ | $^{255}$Es | | | |
| $^{252}$Es | 472 d | α/β$^+$ | $^{248}$Bk/$^{252}$Cf | | | |
| $^{253}$Es | 20 d | α | $^{249}$Bk | | | |
| $^{254}$Es | 276 d | α | $^{250}$Bk | | | |
| $^{255}$Es | 40 d | β | $^{255}$Fm | 20 h | α | $^{251}$Cf |

In order to globally evaluate the buildup of transmutation in other residual MAs, the net and total transmutation rates are calculated for the total concentrations of MAs selected in Table 12. The corresponding values are given in the last two rows in Table 12. The negative net transmutation rate of mixed MAs loading in the FeCrAl cladding signifies that more MAs are found at the EOL than the initial loading. Taking the production of MAs in no MA loading fuels into account, the total transmutation rates are always positive for the six transmutation scenarios. It is noticeable that the total transmutation rates of total MAs in the pure $^{237}$Np loading cases are similar to those of $^{237}$Np in the same case (given in Table 8). The reason is that the concentrations of other MAs generated from $^{237}$Np are rather little compared with the concentration of $^{237}$Np. Due to the production of other MAs from the loaded MAs, the total transmutation rates of total MAs are less than those of the three main loaded MAs (shown in Table 9). From the point of view of the efficiency of transmutation, around 40% (60%) total transmutation rates of investigated MAs for the mixed MAs loading (pure $^{237}$Np loading) cases show that the transmutation method suggested in the present work is an efficient method to reduce the long-lived MAs.

Table 12. Concentrations of MAs at the EOL (in $10^{24}$at./cm$^{-3}$) and the transmutation rates for the total concentration.

| MA | Ref. | FeCrAl | FeCrAl /MAs | FeCrAl /$^{237}$Np | SiC-1 | SiC-1 /MAs | SiC-1 /$^{237}$Np | SiC-2 | SiC-2 /MAs | SiC-2 /$^{237}$Np |
|---|---|---|---|---|---|---|---|---|---|---|
| $^{237}$Np | 1.73E-05 | 1.77E-05 | 2.97E-05 | 4.59E-05 | 1.70E-05 | 4.85E-05 | 9.14E-05 | 1.79E-05 | 4.13E-05 | 7.06E-05 |
| $^{241}$Am | 1.33E-06 | 1.68E-06 | 4.68E-06 | 1.78E-06 | 1.64E-06 | 9.67E-06 | 1.82E-06 | 1.58E-06 | 6.29E-06 | 1.74E-06 |
| $^{243}$Am | 4.84E-06 | 3.25E-06 | 5.79E-06 | 2.88E-06 | 2.79E-06 | 9.42E-06 | 2.81E-06 | 3.58E-06 | 8.73E-06 | 3.61E-06 |
| $^{241}$Cm | 4.31E-14 | 3.30E-14 | 1.59E-13 | 3.21E-14 | 3.05E-14 | 3.59E-13 | 3.23E-14 | 3.63E-14 | 2.90E-13 | 3.83E-14 |
| $^{243}$Cm | 1.57E-08 | 1.24E-08 | 1.61E-07 | 1.17E-08 | 1.10E-08 | 3.78E-07 | 1.14E-08 | 1.30E-08 | 2.89E-07 | 1.35E-08 |
| $^{244}$Cm | 2.23E-06 | 1.27E-06 | 6.89E-06 | 1.10E-06 | 1.04E-06 | 1.30E-05 | 1.04E-06 | 1.44E-06 | 1.17E-05 | 1.44E-06 |
| $^{245}$Cm | 1.59E-07 | 9.93E-08 | 1.05E-06 | 9.08E-08 | 8.14E-08 | 1.99E-06 | 8.62E-08 | 1.09E-07 | 1.59E-06 | 1.15E-07 |
| $^{246}$Cm | 2.57E-08 | 9.56E-09 | 3.29E-07 | 7.39E-09 | 7.15E-09 | 6.46E-07 | 6.79E-09 | 1.21E-08 | 6.40E-07 | 1.16E-08 |
| $^{247}$Cm | 3.98E-10 | 1.37E-10 | 1.15E-08 | 1.08E-10 | 9.93E-11 | 2.35E-08 | 9.81E-11 | 1.73E-10 | 2.30E-08 | 1.72E-10 |
| $^{248}$Cm | 3.54E-11 | 9.93E-12 | 2.21E-09 | 7.56E-12 | 6.87E-12 | 4.53E-09 | 6.76E-12 | 1.31E-11 | 4.79E-09 | 1.29E-11 |
| $^{250}$Cm | 4.02E-18 | 8.84E-19 | 4.00E-16 | 6.76E-19 | 5.72E-19 | 8.17E-16 | 5.90E-19 | 1.22E-18 | 9.36E-16 | 1.24E-18 |
| $^{249}$Cf | 7.39E-14 | 2.48E-14 | 1.56E-11 | 2.08E-14 | 1.74E-14 | 3.38E-11 | 1.89E-14 | 3.05E-14 | 3.05E-11 | 3.28E-14 |
| $^{250}$Cf | 1.82E-13 | 4.16E-14 | 1.85E-11 | 3.06E-14 | 2.72E-14 | 3.83E-11 | 2.69E-14 | 5.75E-14 | 4.22E-11 | 5.63E-14 |
| $^{251}$Cf | 7.73E-14 | 1.88E-14 | 1.13E-11 | 1.42E-14 | 1.22E-14 | 2.36E-11 | 1.25E-14 | 2.51E-14 | 2.46E-11 | 2.56E-14 |
| $^{252}$Cf | 6.30E-14 | 9.74E-15 | 1.01E-11 | 6.38E-15 | 5.83E-15 | 2.06E-11 | 5.40E-15 | 1.48E-14 | 2.68E-11 | 1.39E-14 |
| Total | 2.59E-05 | 2.40E-05 | 4.86E-05 | 5.18E-05 | 2.25E-05 | 8.36E-05 | 9.72E-05 | 2.47E-05 | 7.06E-05 | 7.75E-05 |
| $R_N$(%) | - | - | -12.1 | 34.0 | - | 18.4 | 50.8 | - | 16.1 | 49.0 |
| $R_T$(%) | - | - | 43.2 | 64.5 | - | 40.4 | 62.2 | - | 45.4 | 65.2 |

### 3.4 Comparison with conventional fuel-cladding combinations

The above subsections discussed the critical loadings of MA and the corresponding transmutation efficiencies in candidate ATF combinations. If the same percentages of $^{237}$Np are loaded in the normal UO$_2$-Zircaloy combination, the decrement of the reactivity of the core at EOC is shown in the last column of Table 13. The 2100 pcm negative reactivity shows that 0.31% loading of $^{237}$Np is almost impossible to keep the same cycle length with the same uranium enrichment. 0.60% and 0.78% loading of $^{237}$Np have even much more negative contribution on reactivity. One of the advantages of transmutation in candidate ATF combinations proposed in the present work is that the MAs are transmuted without reducing the reactor cycle length. In other words, the candidate ATF can transmute more MAs with the same cycle length.

Due to the reduction of the cycle length when MAs are added in the normal UO$_2$-Zircaloy case, the transmutation efficiencies at the same EOL burnup level as the candidate ATF combinations (42.5 GWd/t and 46.8 GWd/t for the cladding thickness of 350 μm and 571.5 μm, respectively) are shown in Table 13. Results in Table 8 and Table 13 show that the transmutation efficiencies are almost the same for the normal case and the candidate ATF combinations because the same enrichment of uranium is used. The less than 1% differences of the transmutation efficiency in different cases are due to the competition of neutrons between the cladding and MAs.

Table 13. Transmutation of $^{237}$Np in normal fuel (UO$_2$ or MOX)-cladding (zircaloy) combinations. $R_N$, $R_T$, and $Te$ are the net transmutation rate, total transmutation rate, and the equivalent natural decay time, respectively. $\Delta k_{core}$ is the reactivity between $^{237}$Np loaded core and normal core at the EOC.

| Loading $^{237}$Np | $R_N$ (%) | $R_T$ (%) | $Te$ ($10^6$ year) | $\Delta k_{core}$ |
|---|---|---|---|---|
| 0.31% in UO$_2$ | 42.1 | 65.7 | 3.31 | -0.0210 |
| 0.78% in UO$_2$ | 51.8 | 61.2 | 2.93 | -0.0494 |
| 0.60% in UO$_2$ | 53.1 | 65.3 | 3.28 | -0.0381 |
| 0.14% in MOX | 39.6 | 54.0 | 2.40 | 0.0000 |

Despite of the fact that a typical PWR cannot accept 100% MOX fuel assembly, the LRM is applied to MOX fuel assembly to roughly compare the transmutation of MAs. In the MOX fuel with 9.8% plutonium [33], the Monte Carlo simulations show that only 0.13% mixed MAs and 0.14% $^{237}$Np are allowed to keep the cycle length. The capacity of the transmutation in candidate ATF combinations is much better than the MOX fuel. The last row in Table 13 shows that the transmutation of $^{237}$Np at the EOL (52 GWd/t) is less efficient in the MOX fuel. The transmutation of $^{241,243}$Am in the MOX fuel is also less efficient because they are produced through a set of reactions of plutonium in nuclear reactors. Nevertheless, the transmutation of MAs in MOX fuel may facilitate the post-processing to separate the plutonium and MAs in the spent fuel.

Except for the direct competition of neutrons between MAs and fissile nuclei, the hardening of neutron spectrum plays a role in transmutation because of more fast neutrons in the reactor. The (n,γ) cross sections of MAs generally decrease with the incident neutron energy, while some fission cross sections have sudden increase at fast region, such as those of $^{237}$Np, $^{241}$Am, and $^{243}$Am [28], [35]. The transmutation rates depend on the competitions among these cross sections of MAs and also the competitions between the MAs and dominant fissile nuclei $^{235}$U and $^{239}$Pu. Thus, it is difficult to quantify the general roles of fast neutrons.

**3.5 Relative power and burnup distribution in an assembly**

The relative distribution of power in a fuel assembly is heterogenous because of the different local moderator-to-fuel ratios. Figure 4 displays the relative power distribution of the reference UO$_2$-Zircaloy case at BOL. Higher power is observed in the vicinity of the guide tubes due to the higher local moderator-to-fuel ratio. Peak Factor of Power (PFP) defines the ratio of the maximum power to average power in the fuel assembly. The PFP at the BOL, Middle of Life (MOL, equivalent to 700 EFPDs), and EOL are given in Table 14. The PFPs of the above six transmutation cases are quite similar. The values of the PFP of the transmutation cases are higher than the reference case. However, the differences are within 3% at the BOL and MOL. The maximum discrepancy of the PFP at the EOL is 4%. At the level of the fuel assembly, the loading MAs have little influence on power distribution. The peak value in a core should be further investigated because it is the product of peak factor of assemblies in core, peak factor of pins in assembly (results in Table 14), and axial peak factor.

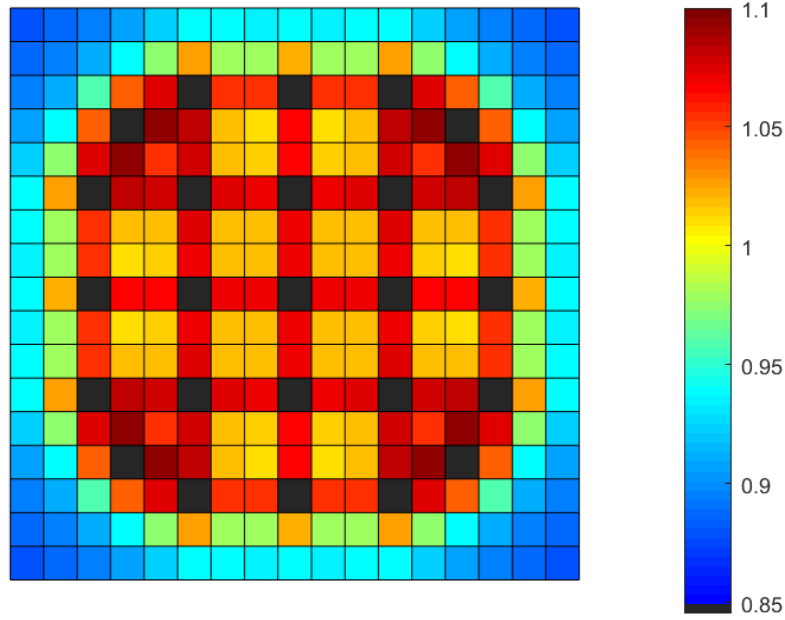

Figure 4. Relative power distribution at the BOL for the reference $UO_2$-Zircaloy case. The deep blue points out the guide tubes.

Table 14. Peak Factor of Power (PFP) at the BOL, MOL, EOL and Peak/Valley Factor of Burnup (PFB/VFB) at the EOL for the reference $UO_2$-Zircaloy case and six transmutation cases: FeCrAl cladding, 350 μm (SiC-1) and 571.5 μm (SiC-2) thickness SiC cladding with mixed MAs and $^{237}$Np loading

| Case | PFP BOL | PFP MOL | PFP EOL | PFB EOL | VFB EOL |
|---|---|---|---|---|---|
| Reference | 1.09 | 1.06 | 1.02 | 1.06 | 0.91 |
| FeCrAl/MAs | 1.11 | 1.08 | 1.06 | 1.08 | 0.90 |
| FeCrAl/$^{237}$Np | 1.10 | 1.08 | 1.05 | 1.08 | 0.90 |
| SiC-1/MAs | 1.11 | 1.09 | 1.06 | 1.08 | 0.89 |
| SiC-1/$^{237}$Np | 1.11 | 1.09 | 1.06 | 1.08 | 0.90 |
| SiC-2/MAs | 1.10 | 1.08 | 1.04 | 1.08 | 0.90 |
| SiC-2/$^{237}$Np | 1.10 | 1.08 | 1.04 | 1.07 | 0.90 |

Burnup is the physical quantity to measure the accumulated power. The burnup at the EOL reflects the utilization of fuel. Higher (lower) burnup at the EOL means more (less) efficient utilization of fuel. The Peak Factor of Burnup (PFB) and the Valley Factor of Burnup (VFB) are defined to present the burnup at the EOL in a fuel assembly. Figure 5 exhibits the relative burnup distribution of the reference case at the EOL. Higher burnup is observed in the vicinity of the guide tubes due to the higher power induced by the higher local moderator-to-fuel ratio. The corresponding PFB and VFB for the reference case and the above six transmutation cases are given in the last two columns in Table 14. Because the burnup is proportional to the time-accumulated power, the PFB at the EOL for each case is within the corresponding interval [PFP_EOL, PFP_BOL]. Similar PFB and VFB values are shown among the six transmutation cases determined in the present work. About 2% difference exists between the reference and the transmutation cases for the PFB at the EOL. Nevertheless, only 1% difference is observed for the VFB at the EOL. Therefore, the loading MAs has little influence on the efficiency of the utilization of fuel.

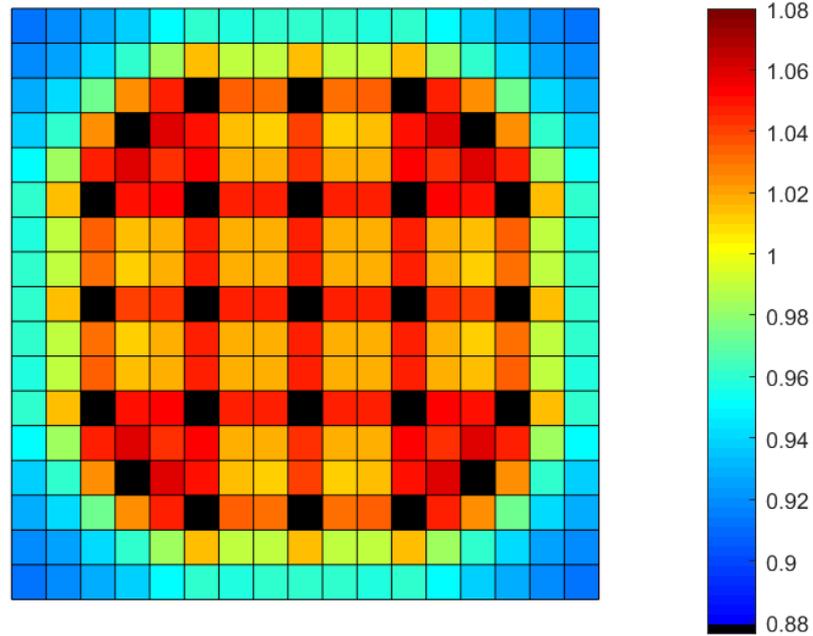

Figure 5. Relative burnup distribution at the EOL for the reference UO$_2$-Zircaloy case. The deep blue points out the guide tubes.

### 3.6 Radial power distributions

As new fuel-cladding combinations, investigations on their radial distribution of power are of importance because the radial power distribution can give feedback to neutronic calculations and multi-physics coupling study. The Monte Carlo simulations performed with the geometry shown in Figure 2 reveal the similarity of the radial power distribution as a function of the effective full power depletion time for the above six transmutation cases. Figure 6 shows the radial relative power distributions for the FeCrAl cladding and SiC-1 cladding cases at the BOL, MOL, and EOL. As defined in Section 2.3, $x$ represents the relative radius, such as $x=0$ ($x=1$) stands for the center (outer surface) of the fuel pellet. The periphery phenomenon is induced by the radial distribution of thermal neutron flux and the radial distribution of isotopic concentrations after the BOL. The corresponding explanations, descriptions, and influence have been detailed in Refs. [2], [21], [34].

On the other hand, an empirical analytical formula is shown to nicely describe the radial power distribution for the UO$_2$-zircaloy system, three ATF U$_3$Si$_2$-FeCrAl cases, and the UO$_2$/U$_3$Si$_2$-FeCrAl candidate ATF [34]. The coefficients and corresponding uncertainties of the analytical formula (such as Eq. (5) in Section 2.3) at different positions are given in Table 16 in the appendix. Figure 7 shows the ratios of relative power for the FeCrAl-MA case transmutation case to the empirical formulae as a function of fuel exposure for several radial positions. Figure 8 illustrates the ratios of the power calculated by RMC code for transmutation fuels to the empirical formula versus the relative radial position at BOL, 500 EFPDs, and EOL, respectively.

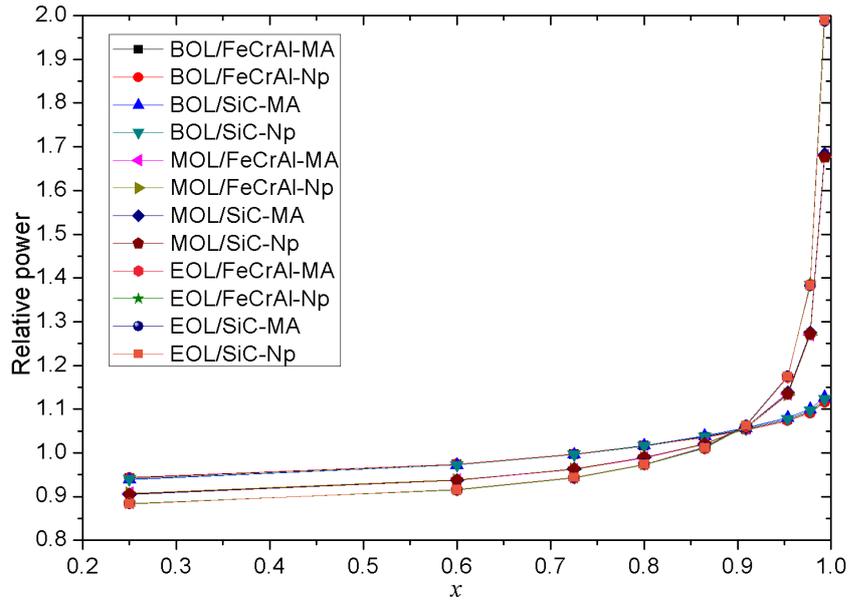

Figure 6. Radial power distribution for 350 μm thickness FeCrAl and SiC cladding with mixed MAs loading and $^{237}$Np loading at the BOL, MOL, and EOL. No evident difference is observed among different cases.

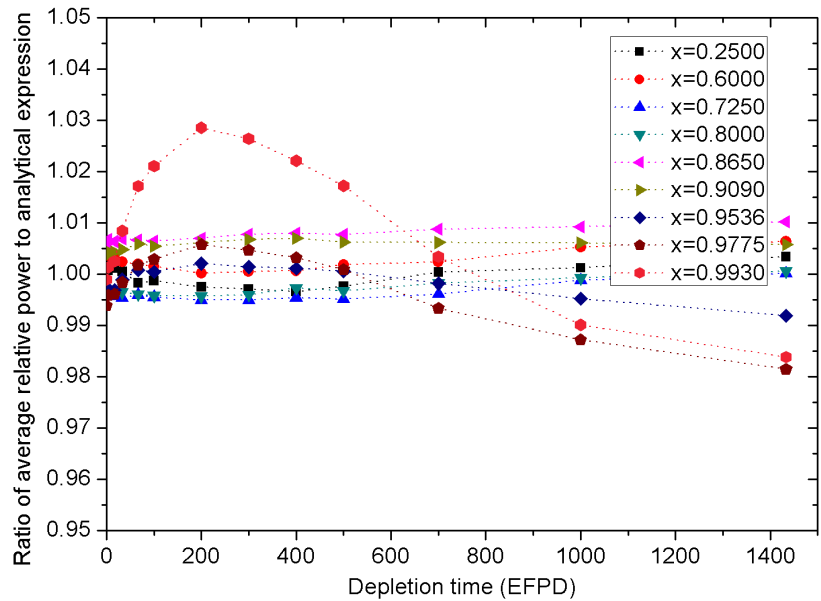

Figure 7. Ratios of relative power of mixed MAs loaded FeCrAl cladding to the empirical formulae at different relative radial positions (x). The discrepancies of radial power distribution between other transmutation fuels and the FeCrAl-MA case are within 1.0%.

As shown in Figure 7 and Figure 8, the Monte Carlo simulation results show almost the same radial relative power distributions for the six transmutation cases. In addition, the radial power distribution of the transmutation fuel rods is similar to the current UO$_2$-zircaloy system, three candidate ATF U$_3$Si$_2$-FeCrAl cases, and one UO$_2$/U$_3$Si$_2$-FeCrAl combination [34]. The relative differences of the radial power distribution between transmutation cases and empirical formula proposed in the previous work are less than 3%. In the region $x < 0.95$, the relative differences are within 1%. The difference of burnup distribution between the transmutation fuel rods and the normal case is within 2%.

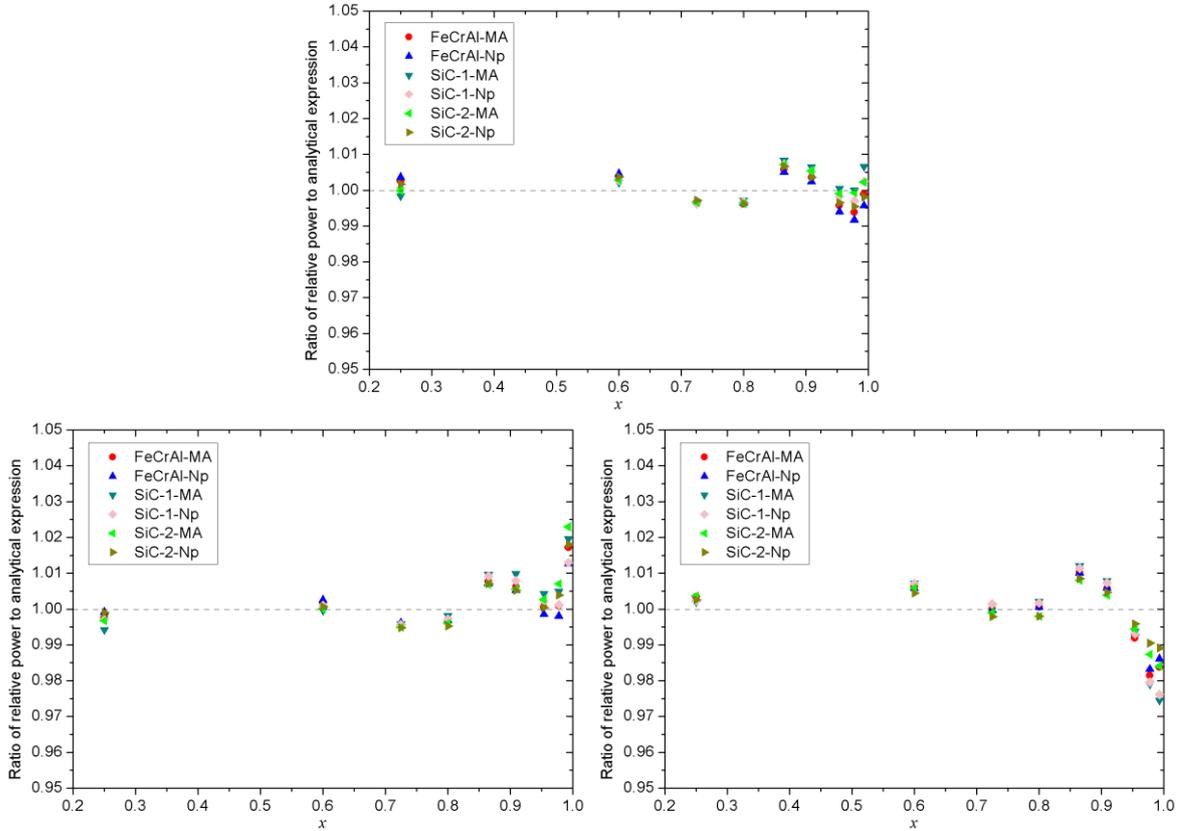

Figure 8. Ratios of power calculated with RMC to empirical formulae at the BOL, at 500 EFPDs, and at the EOL. FeCrAl and SiC-1 (SiC-2) stand for FeCrAl cladding and 350 μm (571.5 μm) thickness SiC cladding, respectively. MA (Np) represents the critical mixed MAs ($^{237}$Np) loading.

Table 15. Thermal neutron capture cross sections (in barn) in ENDF/B-VII.0 and JEFF-3.1.1

|  | $^{238}$U | $^{237}$Np | $^{241}$Am | $^{243}$Am | $^{244}$Cm | $^{245}$Cm |
|---|---|---|---|---|---|---|
| ENDF/B-VII.0 [28] | 2.68 | 162 | 618 | 75.1 | 15.1 | 359 |
| JEFF-3.1.1 [35] | 2.68 | 181 | 647 | 76.7 | 10.4 | 359 |

After BOL, the cases of transmutation have a little sharper power distribution than the normal cases due to the better breeder ability of MAs than $^{238}$U. As the data shown in Table 15, the thermal neutron capture cross sections of MAs are quite larger than that of $^{238}$U in two evaluated nuclear data libraries. The concentrations of MAs decrease with the radius due to the spatial distribution of the thermal neutrons in the fuel rod. A direct consequence is that the positive contribution of MAs on the thermal power becomes less important after about 300 EFPDs. In addition, higher fission rate leads to more accumulated Fission Products (FPs), which have a negative contribution to the reactivity. After the MOL, the radial distribution of power in the cases of transmutation is a little lower than that in the normal cases (Figure 7) because the negative contribution to the reactivity induced by the accumulated FPs becomes more important than the positive contribution of MAs. In fact, at high burnup level, a higher concentration of MAs near the surface signifies higher production rates than consummation rates of MAs, which implies that the contribution of MAs on the power becomes almost the same as the normal cases.

### 3.7 Radial distribution of isotopic concentrations

The investigations on the radial distribution of the isotopic concentrations of the major actinides, such as uranium and plutonium, provide information to neutronic calculation. For example, the radial power distribution is based on both the neutron spectrum and the isotopic concentration of the fissile isotopes. In the present studies, it is assumed that there is no mass transfer in the fuel. The principal fissile isotopes in a reactor are $^{235}$U, $^{239}$Pu, and $^{241}$Pu in the uranium fuels. The radial concentration distributions of the major actinides for the above six transmutation cases are similar. The ratios of the relative concentrations of the major actinides in the transmutation fuels to the five normal cases (the current fuel-cladding system and four candidate ATF combinations) are shown in Figure 9. The BOL is 3.3 EFPDs in the treatment of the concentrations of $^{239}$Pu and $^{241}$Pu because there is no plutonium in the fuels at the BOL.

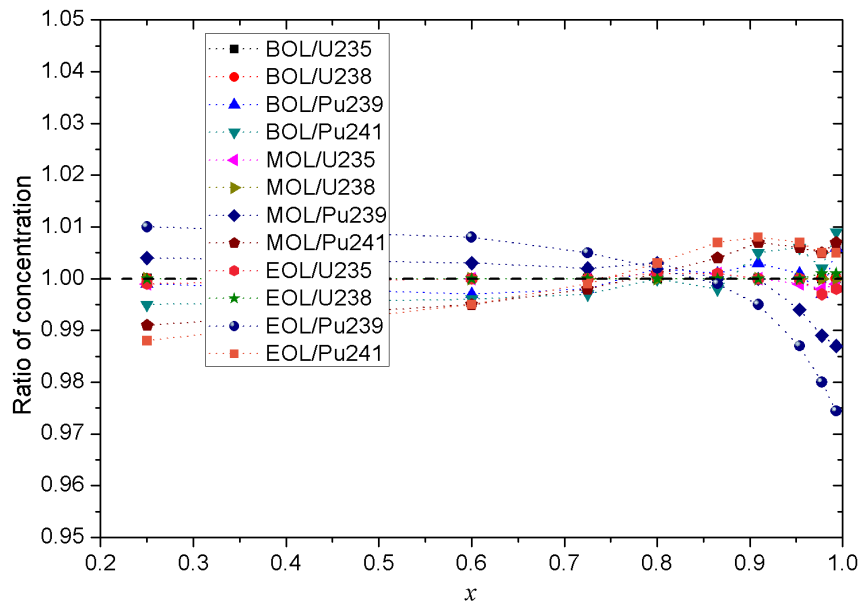

Figure 9. Ratios of average major actinide concentration in transmutation fuels to normal fuels at the BOL, MOL, and EOL. The dispersions to each case are within 1% for the above four major actinides.

Another important phenomenon is that the additional MAs in the U$_3$Si$_2$ fuel has almost no influence on the radial distribution of concentrations of the major actinides ($^{235}$U, $^{238}$U, $^{239}$Pu, and $^{241}$Pu) for the six transmutation designs proposed in the present work (Figure 9). About 2.5% lower $^{239}$Pu concentration at the periphery in the MAs loaded fuels is due to the competition between the neutron capture reaction of MAs and $^{238}$U.

## 4. Conclusions

The U$_3$Si$_2$ fuel is an excellent candidate ATF due to its high thermal conductivity and high uranium density. The stainless steel FeCrAl and the ceramic material SiC are treated as the potential ATF cladding materials due to their excellent oxidation resistance. Previous investigations showed the satisfactory neutron economy of $U_3Si_2$ based candidate ATF-Cladding combinations [2], [9]. On the other hand, the transmutation is an efficient method to reduce long-lived MAs, which have a negative contribution to the reactivity. The present study performs on the transmutation of MAs in the 4.9% enriched candidate ATF combinations $U_3Si_2$-FeCrAl and $U_3Si_2$-SiC.

Based on the cycle length analysis by using LRM, the critical loading concentrations of $^{237}$Np and the five mixed MAs loadings in the $U_3Si_2$ fuel are determined for 350 μm thickness FeCrAl cladding and 350 μm and 571.5 μm thickness SiC claddings. The Monte Carlo simulations show the excellent transmutation efficiency for MAs. For example, the total transmutation rates of the longest-lived MA $^{237}$Np in the above six cases are equivalent to $3\times10^6$ years of natural decay time. If only $^{237}$Np is considered, one $U_3Si_2$-SiC assembly can transmute $^{237}$Np produced from six normal assemblies. In the mixed MAs transmutation cases, the net (total) transmutation rates are about (more than) 90% for $^{241}$Am, which is the richest MA in the spent fuel considered in this study. The results show that the transmutation can be also an efficient method to reduce the quantity of $^{243}$Am. The loadings of pure $^{237}$Np lead to more production of plutonium, which can be recycled to fabricate MOX fuel. The negative contribution of pure $^{237}$Np loading on other MAs is an advantage of this kind of design. The mixed MAs loading induces higher concentration of MAs. According to the analyses on total transmutation rate of MAs with half-lives longer than 1 year, the transmutation is efficient to globally reduce the MAs.

Quite similar distribution of the power and burnup in a fuel assembly is observed for the six transmutation cases proposed in the present work. The loading MAs have small influence on the power distribution in an assembly. Further investigations are required to calculate the peak value of power in a core because it is the product of three peak factors, including the peak factor in an assembly investigated in the present work. Nevertheless, the transmutation cases almost do not change the efficiency of the fuel utilization from the point of view of the relative minimum burnup at EOL. The loading of MAs may have impact on core safety parameters, including the Doppler coefficient, the boron efficiency, and the effective delay neutron. More detailed investigations on these properties should be further studied before application.

The loadings of MAs proposed in the present work have slight influence on the radial distribution of the power and isotopic concentrations of the major actinides. Therefore, the additional MAs do not change the distribution of the local power and burnup in the fuel. The competition of the reactions with the neutron between MAs and major actinides has less than 3% (1%) influence on the radial distributions of power and concentration(s) of $^{239}$Pu ($^{235}$U, $^{238}$U, and $^{241}$Pu) during the serving life of the fuel assemblies.

It should be noted that in general simulation results have certain uncertainties from the numerical methods, simulation model, nuclear data libraries, and so on. Many methods can be used to investigate the theoretical uncertainties. For example, the recent suggested uncertainty decomposition method is used to decompose the systematic and statistical uncertainties of a simple nuclear mass model [36]. The uncertainties from the numerical methods (Monte Carlo here) and the nuclear data libraries (ENDF/B-VII.0 and JEFF-3.3) are shown to be not significant for the transmutation study [16].

## Acknowledgments

The authors acknowledge the authorized usage of the RMC code from Tsinghua University for this study. This work has been supported by the National Natural Science Foundation of China under Grant No. 11775316, the Tip-top Scientific and Technical Innovative Youth Talents of Guangdong special support program under Grant No. 2016TQ03N575, and the Fundamental Research Funds for the Central Universities under Grant No. 17lgzd34.

## Appendix

Table 16 Coefficients and associated uncertainties and the coefficient of determination for fitting at each position for relative power distribution obtained with the average of current $UO_2$-zircaloy and four candidate ATF combinations [34]

| $x$ | $10^9 a(x)$ | | $10^5 b(x)$ | | $c(x)$ | |
|---|---|---|---|---|---|---|
| 0.2500 | 7.47 | 19.4% | -5.18 | 5.19% | 0.94 | 0.08% |
| 0.6000 | 9.79 | 9.30% | -5.63 | 2.98% | 0.97 | 0.05% |
| 0.7250 | 11.88 | 7.00% | -5.71 | 2.69% | 1.00 | 0.04% |
| 0.8000 | 12.07 | 4.50% | -5.07 | 1.98% | 1.02 | 0.03% |
| 0.8650 | 10.12 | 7.20% | -3.52 | 3.86% | 1.03 | 0.04% |
| 0.9090 | 5.46 | 13.0% | -0.44 | 29.8% | 1.05 | 0.04% |
| 0.9536 | -10.96 | 17.4% | 8.77 | 4.03% | 1.08 | 0.09% |
| 0.9775 | -51.54 | 8.70% | 29.12 | 2.87% | 1.10 | 0.21% |
| 0.9930 | -205.6 | 6.10% | 93.54 | 2.49% | 1.12 | 0.58% |